\documentclass{PoS}

\title{First prototype of a silicon tracker using an 'artificial retina' for fast track finding}

\ShortTitle{First prototype of silicon tracker with 'artificial retina'}

\author{\speaker{N.~Neri},
A.~Abba, F.~Caponio, M.~Citterio, S.~Coelli, J.~Fu, A.~Geraci, M.~Monti, M.~Petruzzo \\
        INFN, Sezione di Milano, and Politecnico di Milano (Italy)\\
        E-mail: \email{nicola.neri@mi.infn.it}}
\author{F.~Bedeschi, P.~Marino, M.~J.~Morello, A.~Piucci, G.~Punzi, F.~Spinella, S.~Stracka, J.~Walsh\\  INFN, Sezione di Pisa, Universit\`a di Pisa, 
 and Scuola Normale Superiore di Pisa (Italy)}
\author{L.~Ristori\\ INFN, Sezione di Pisa (Italy), and Fermilab (USA)}
\author{D.~Tonelli\\ CERN (Switzerland)}

\abstract{We report on the R\&D for a first prototype of a silicon tracker 
based on an alternative approach for fast track finding. 
The working principle is inspired from neurobiology, in particular by the processing of 
visual images by the brain as it happens in nature. It is based on extensive parallelisation 
of data distribution and pattern recognition. 
In this work we present  the design of a practical device that consists of a telescope based on 
single-sided silicon detectors; we describe the data acquisition system and the implementation 
of the track finding algorithms using available digital logic of commercial FPGA devices. 
Tracking performance and trigger capabilities of the device 
are discussed along with perspectives for future applications. }

\FullConference{Technology and Instrumentation in Particle Physics 2014,\\
		2-6 June, 2014\\
		Amsterdam, the Netherlands}

\begin{document}

\section{Introduction}

The low level trigger capabilities of tracking systems have been proved to be crucial for the success 
 of high energy physics experiments, especially at hadron colliders. 
The ``artificial retina'' is an approach for fast track finding based on an 
algorithm~\cite{Ristori:2000vg} that takes inspiration from neurobiology,
in particular by the fact that specific neurons of the retina are specialised to identify specific shapes.
In the case of the ``artificial retina'' we construct a grid of receptors that are tuned to identify charged 
particle tracks  with different parameters. In this work we present the design of the first prototype 
of a tracking system with ``artificial retina'' capable 
of fast track finding  with a latency $<1~\mu$s and with track parameter resolutions 
that are comparable with the offline results. The prototype detector has 
 an active area of about 100 cm$^2$ and the maximal rate for track 
reconstruction is about 1 MHz. 
However, the ``artificial retina'' is a modular system that can be designed to work 
for HEP applications, {\it i.e.} high rates and large detectors, providing offline-like 
track quality results with a sub-$\mu$s latency~\cite{Abba:2014lhcb, Punzi:2014, Tonelli:2014}.
%
%
\section{The Retina Algorithm}
\label{sec:retina_alg}

\begin{figure}[!h]
\centering
\includegraphics[width=0.9\textwidth]{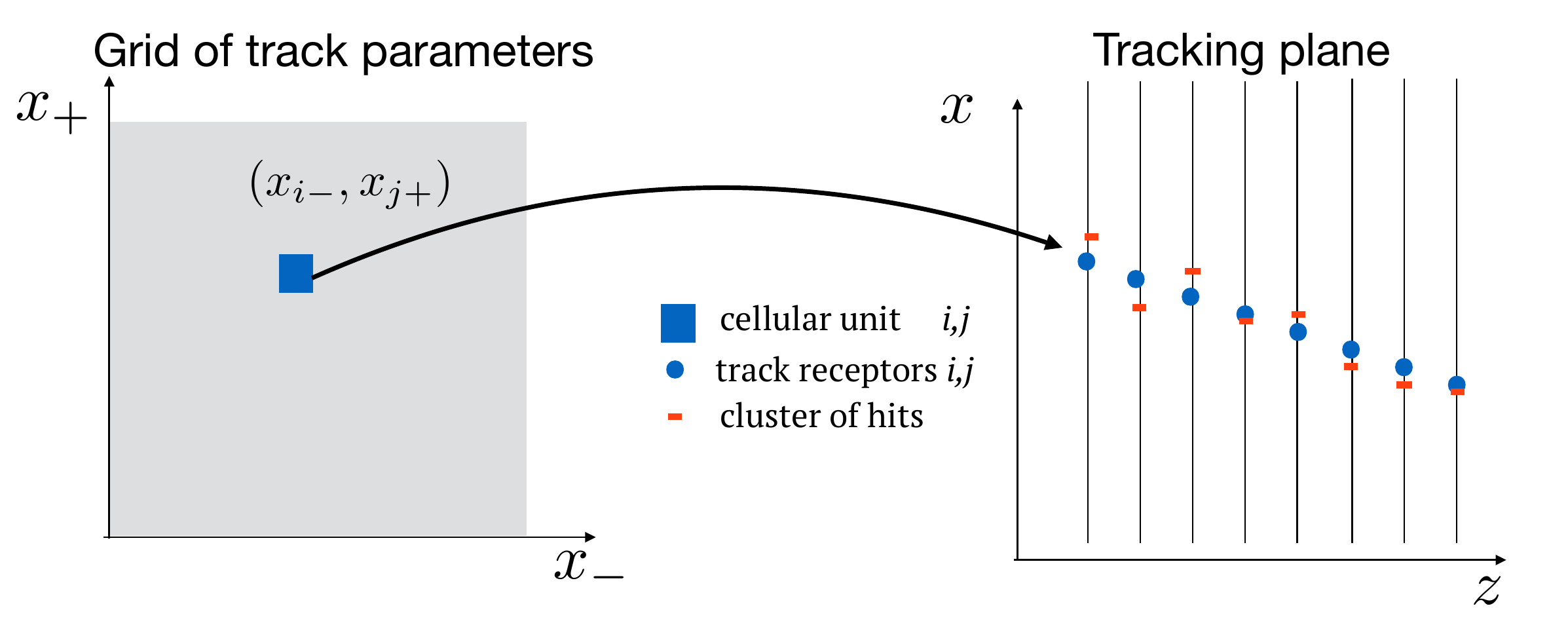}
\caption{A cellular unit in the grid of parameters ($x_-,x_+$) identifies
 a specific track hypothesis. The corresponding track receptors represent 
 the intercepts of the ideal track with the tracking layers. Cluster of hits corresponding 
 to matched tracks are aligned with the track receptors within the experimental resolution.}
\label{fig:Cell_TrackingPlane}
\end{figure}

\begin{figure}[!h]
\centering
\includegraphics[width=0.6\textwidth]{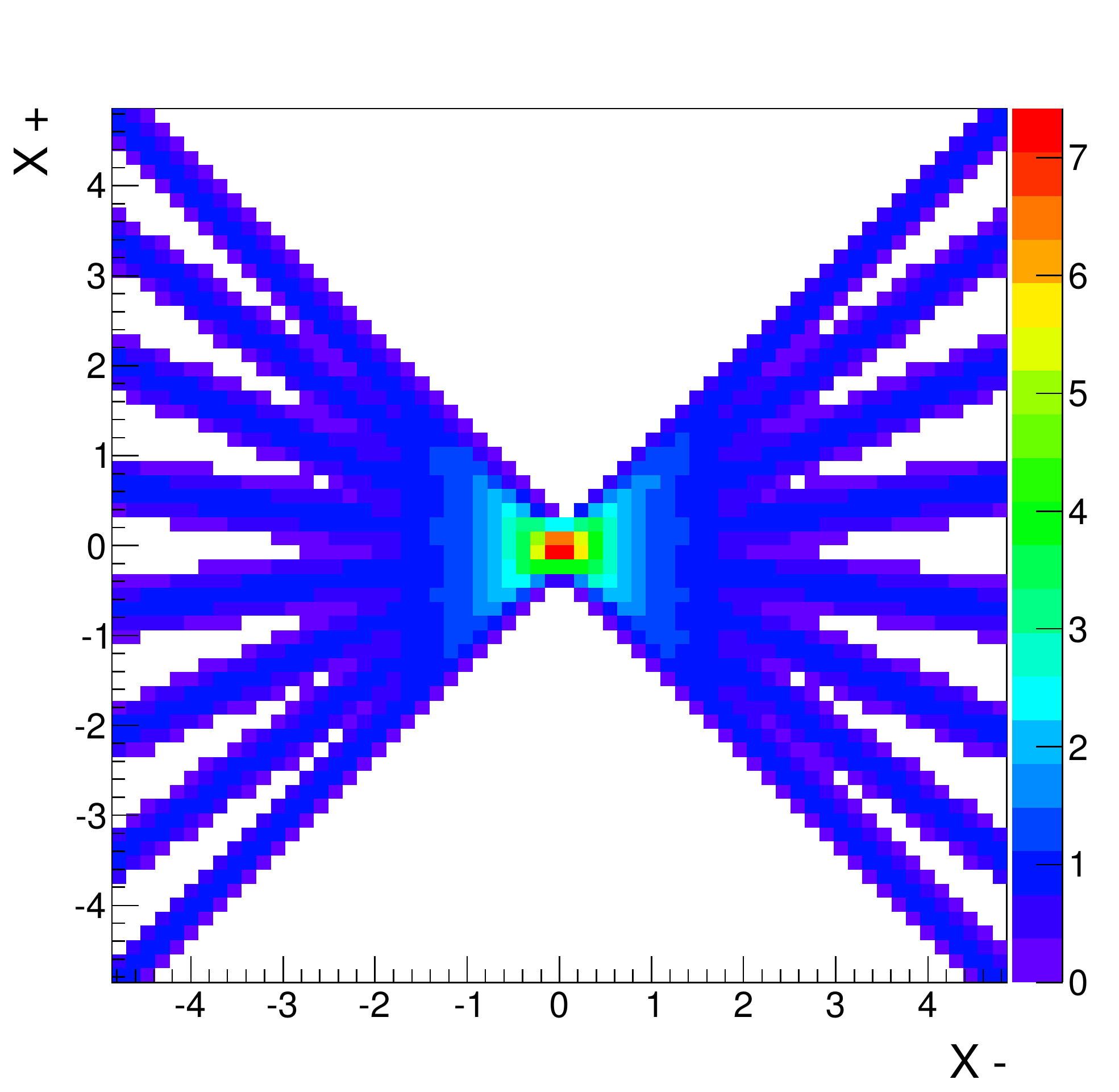}
\caption{Simulation of the ``artificial retina'' response for the prototype silicon tracker with 
 8 layers. The intensity of the response (weight) is 
  indicated by the colour code reported on the right. A reconstructed track is identified by a local 
maximum in the  track parameter space.}
\label{fig:track}
\end{figure}

For simplicity's sake, let's consider the reconstruction of a charged particle track 
 in the 2D space ($x, z$) using a tracking system composed of multiple layers.
If ($x_f, z_f$) and  ($x_l, z_l$) are the coordinates of a cluster on the first 
 and the last tracking detectors, respectively, and  $x_{\pm} =(x_f\pm x_l)/2$ 
and $z_\pm = (z_f \pm z_l)/2$, we can define the equation of a 2D track as 
\begin{equation}
\label{eq:2d_trk}
x(z) = x_+ + x_- (z-z_+)/z_-.  
\end{equation}
The grid of track parameters ($x_-,x_+$) is shown in Fig.~\ref{fig:Cell_TrackingPlane},
 where a cellular unit at position $ij$ corresponds to track receptors representing the intercepts of 
 the ideal track with the detectors. A Gaussian receptor field 
 is used to evaluate the goodness of the match of a set of clusters with a specific track hypothesis
and a weight, $W_{ij}$, for each cell is calculated as
\begin{eqnarray}
\label{eq:weight}
W_{ij} = \sum_k \exp{\left(-\frac{s_{ijk}^2}{2\sigma^2}\right)} &\textrm{\quad if }& s_{ijk}<2\sigma,  \\
W_{ij} = 0 \hspace{2.5cm} &\textrm{\quad if }& s_{ijk}>2\sigma, \nonumber
\end{eqnarray}
where $s_{ijk}= |x -x_{j,+} - x_{i,-} (z_k-z_+)/z_-|$  is the distance of the cluster 
from the intercept of the track, associated with that cell, in layer $k$. 
The width of the receptor response field, $\sigma$, is much larger that the 
 obtainable resolution on the track parameters and has to be adjusted for optimal response. 
A reasonable choice is $\sigma \simeq \Delta$, where $\Delta$ is the granularity of 
the grid of track parameters.
A cluster ($x,z$) is represented in the ($x_-,x_+$) plane as a line,
\begin{equation}
\label{eq:cluster_line}
x_+ = - x_- \frac{z-z_+}{z_-}+x.
\end{equation}
%
The cluster information is sent in parallel to appropriate cellular units that 
 calculate the weights. A matched track would result in a local maximum of 
the retina response $W_{ij}$, as reported in Fig.~\ref{fig:track}.

\section{Tracking System Prototype with ``Artificial Retina''}
\label{sec:telescope}
\begin{figure}[!h]
\centering
\includegraphics[width=0.6\textwidth]{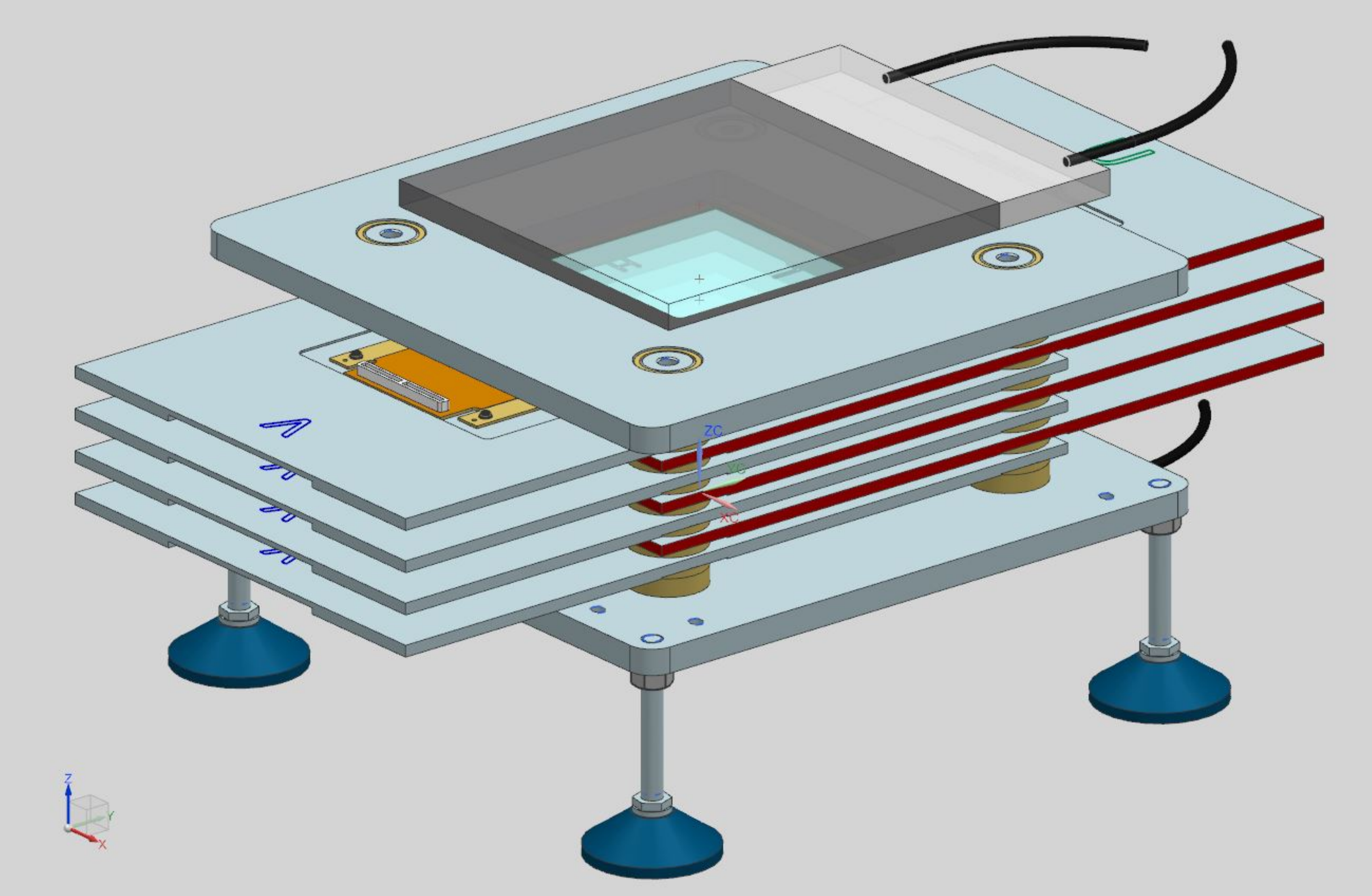}
\caption{Telescope layout. The telescope consists of 8 planes of single-sided silicon sensors. 
Plastic scintillators are placed on the top and the bottom 
of the telescope providing external trigger for charged particle tracks traversing the tracking volume.}
\label{fig:Telescope}
\end{figure}

The prototype system consists of 8 planes of single-sided silicon sensors with 512 strips each 
and 183 $\mu$m pitch; the  active area of the sensor is about 100 cm$^2$ with 500 $\mu$m thickness. 
 The telescope layout is shown in Fig.~\ref{fig:Telescope}. 
 The sensors are arranged at a distance of 0.8 cm from each other 
 and plastic scintillators are positioned at the top and the bottom of the telescope providing 
trigger signals for charged particle tracks traversing the tracking volume.
\par
The expected rate for cosmic rays at sea level traversing all the telescope layers is of the order of 1 Hz.
 The test with cosmic rays allows to verify all the functionalities of the system and to prove the working principle. 
 The sensors are readout by a custom  ASIC~\cite{Lochner:2006vba} which can accept trigger rates up to 1.1 MHz 
and eventually a test with beam is envisaged to verify the behaviour at higher rates.

\begin{figure}[!h]
\centering
\includegraphics[width=0.9\textwidth]{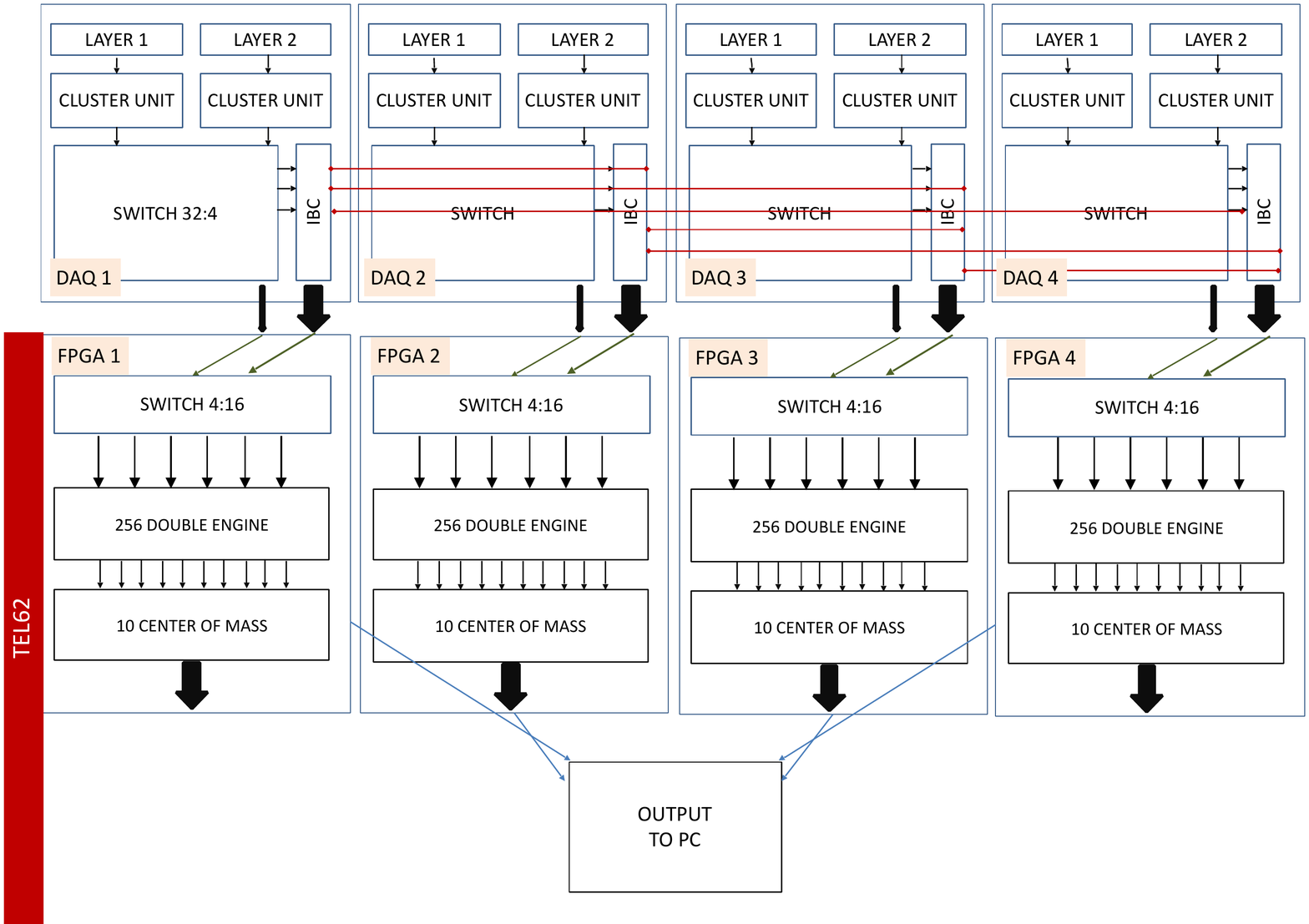}
\caption{Schematics of the telescope readout system and of the retina architecture based on a TEL62 board.}
\label{fig:retina_architecture}
\end{figure}

The schematic of the telescope readout system is shown in Fig.~\ref{fig:retina_architecture}. 
It consists of a custom data acquisition (DAQ) board  described in detail in Ref.~\cite{Caponio:2014rt};
 here we report a short description of it.
The ASIC provides the measurement of the hit position and the pulse height of 128 channels; 
 each detector is readout by 4 ASICs and each one multiplexes 32 channels on a single analog output, 
for a total of 16 analog output channels for the readout of a plane of the telescope. 
 The DAQ boards are based on FPGAs Xilinx Kintex 7 lx160, each one connected to 4 ADCs with 8 serial outputs at 
80 mega-samples per second; they manage the readout ASICs and the sampling of the analog channels. 
The readout is performed at 40 MHz on 4 channels for each ASIC that corresponds to a decoding 
of the telescope information at 1.1 MHz.

A second stage of the signal processing, for the calculation of the retina response, resides in the 
TEL62 board~\cite{Angelucci:2011jma}.  
 The ``artificial retina'' has been implemented using commercial FPGAs and its design is based on three main logic modules: 
 a switch for the detector hits, a pool of engines for the digital processing of the hits, 
 and a block for the calculation of the track parameters. 
 The TEL62 board is equipped with 4 Altera Stratix III FPGAs that provide adequate computing performance 
 for the switch, 
 the engines, and the interpolation for determining the track parameters. 
 The parameters of the tracks detected are finally transferred to host PC via Ethernet lane at 4 Gbit/s. 
 The FPGA resources of the TEL62 boards have been divided among the different modules: 
 approximately 13\% for the switch module, 50\% for the pool of engines and 12\% for the track parameter determination. 
 The rest is kept for backup.
 This configuration allows to realise an ``artificial retina'' with about 2000 cells with a clock  
 frequency of the system of about 200 MHz. In general the number of cells depends on the 
requested precision on track parameters and their range. 
The number of cells scales linearly with the FPGA resources and the system is modular.
The latency of the ``artificial retina'' response is determined by the number of 
 clock cycles necessary for the hits to pass through the entire system, which is estimated to be less than 100.
At a clock frequency of about 200 MHz the latency is below 1 $\mu$s and the system would allow for fast trigger 
 decisions based on real time track parameter determination.
The effect of the occupancy on the latency of the retina response is negligible for track rates 
below 1 MHz and the latency is constant.

\subsection{The Switch Module}
The grid of the retina parameters is divided in 4 regions corresponding to the number of available FPGAs in the 
TEL62 board. The switch module distributes in parallel the hits from all the 8 detector layers to appropriate 
logical units,  called engines, which determine the good match (weight) of the hit combinations 
with a specific track hypothesis. 
A cluster is represented as a line in the space of parameters,
\begin{equation}
\label{eq:switch_line}
\left|x-x_+-x_-\frac{z-z_+}{z_-}\right|<2\sigma
\end{equation}
 and its information has to be delivered to the engines belonging to the regions defined by Eq.~\ref{eq:switch_line},
 that reside in different FPGAs.
  The path of each hit is determined according to the cluster coordinates by a set of look-up tables (LUTs) 
and it is finally distributed to a set of  engines corresponding to track parameters that are compatible 
with that specific hit. 
A graphical representation of the switch module is shown in Fig.~\ref{fig:retina_architecture}, which is 
organised in two  levels.
The first level of the switch is placed in the Kintex-7 FPGA 
of the readout board and distributes data from 32 inputs to 4 outputs. 
The 8 layer telescope is readout using 4 DAQ boards: 
each readout board receives signals from 2 detector planes on 32 analog inputs. The switch  module
 sends the signals on 4 output ports: one output port is connected to the second level of the switch 
in the TEL62 board, and the other output ports are connected to the other three readout boards for 
 a fully meshed network. The second level of the switch resides in the FPGAs of the TEL62 board and distributes  the hits from 
 4 inputs to 16 outputs,  each one delivering the data in parallel to 16 engines.
 The switch module relies on a basic $16\times16$ way data dispatcher where only specific input and output ports are connected.
 Each engine calculates the weight for 2 cellular units for a total of 2048 cellular units for the retina.

\subsection{The Engine Module}
 The engine corresponds to a specific cell at position $ij$
 in the grid of the track parameter space ($x_{-}, x_{+}$) 
 and its excitation is proportional to the weight, $W_{ij}$, defined in Eq.~\ref{eq:weight_time}.
\begin{figure}[!h]
\centering
\includegraphics[width=0.6\textwidth]{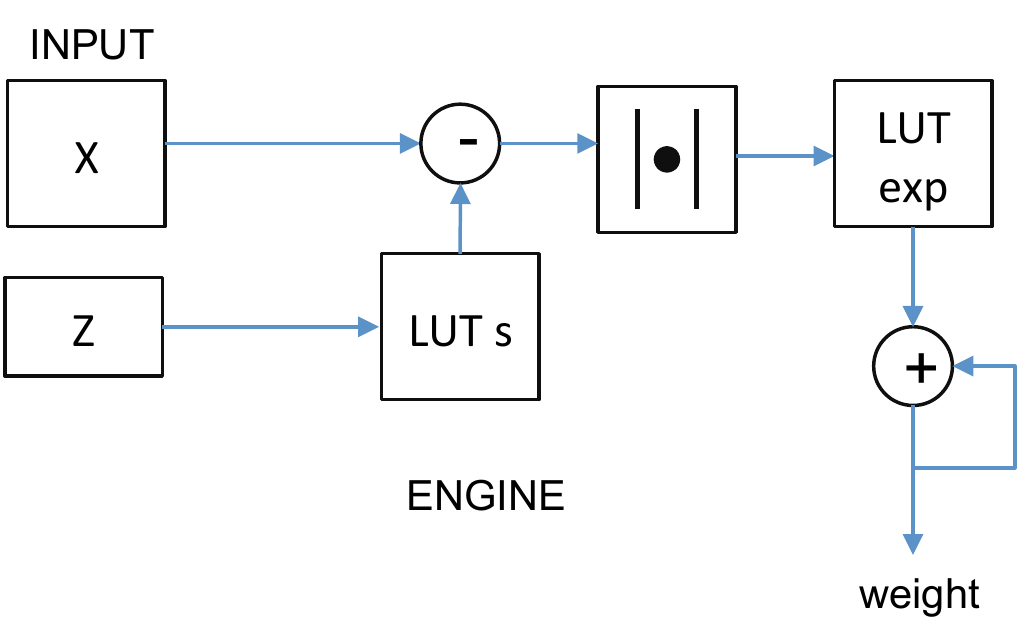}
\caption{Diagram of the logical blocks of the engine.
The distance between the cluster with coordinates ($x, z$)  from the track intercept  
 in a specific layer is calculated using a LUT. Another LUT is used for the evaluation of the 
 exponential function of the weight.}
\label{fig:engine}
\end{figure}
  In Fig.~\ref{fig:engine} it is shown the diagram of the logical blocks of the engine where
 two LUTs with 1024 16-bit integers have been used for the calculation of the weights.  
One LUT is used for the calculation of the distance $s_{ijk}$ and the other LUT is used for the calculation 
 of the exponential function.
Each engine calculates the weights for two neighbouring cells in the retina of the track parameters. 
Each engine interacts with the nearest neighbour engines to determine the maximal weight 
 value and then another logical block interpolates the track parameters from other two adjacent 
engines in the appropriate directions.

\subsection{The Track Module}

\begin{figure}[!h]
\centering
\includegraphics[width=0.9\textwidth]{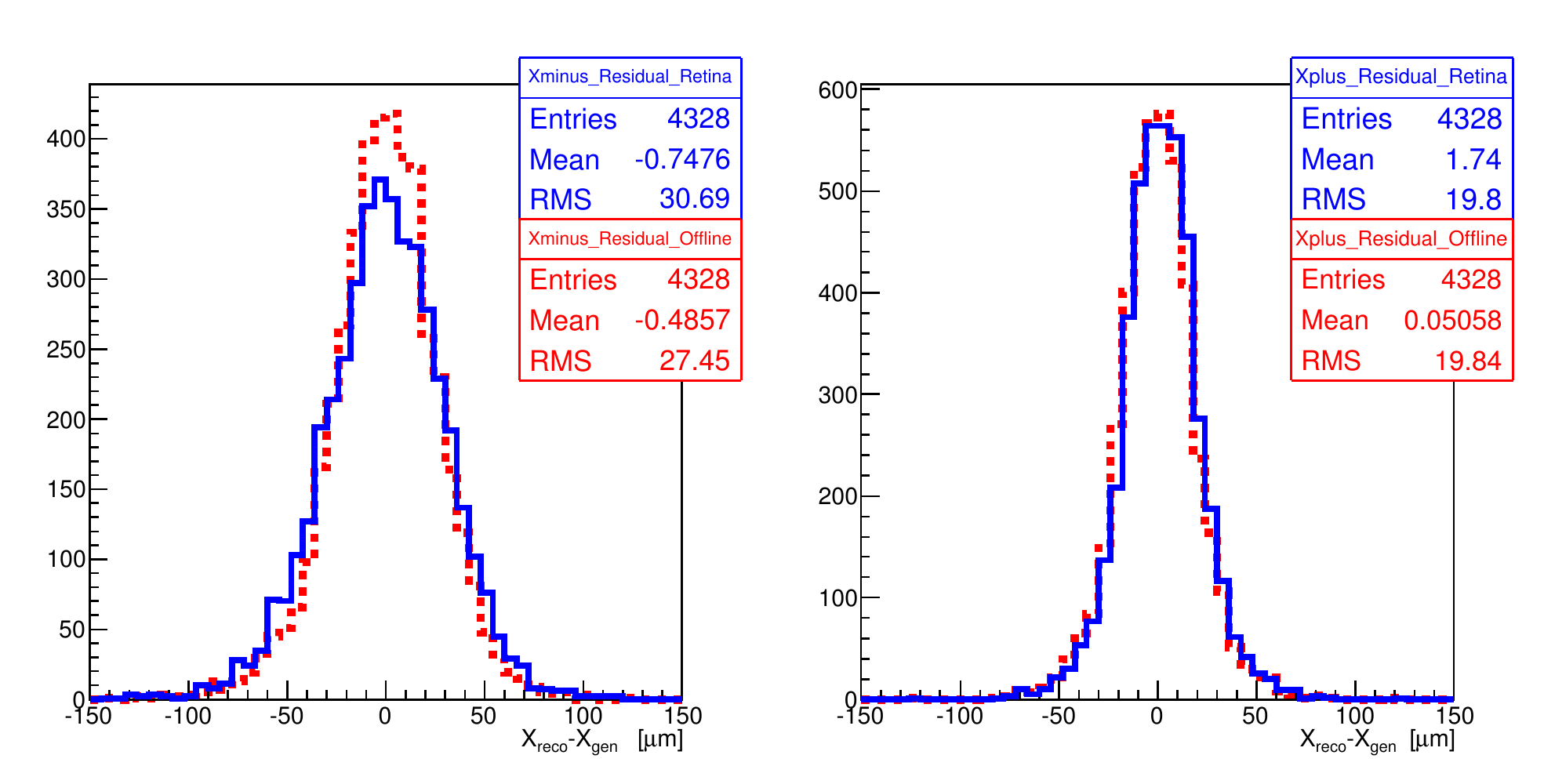}
\caption{Resolution for $x_-$ (left) and for $x_+$ (right) track parameters obtained 
with the ``artificial retina'' (open blue histogram) compared with offline results (dashed red histogram).}
\label{fig:resolution}
\end{figure}

%
%
The engine determines if the calculated weight corresponds to a local maximum and then
 sends its weight, and the ones from the adjacent cells, to the track module.
The track parameters are determined by Gaussian interpolation of the weights of neighbouring cells
\begin{equation}
\label{eq:Gauss_interpol}
x = x_0+\frac{\Delta}{2}\frac{\log(W_-/W_0)-\log(W_+/W_0)}{\log(W_-/W_0)+\log(W_+/W_0)},
\end{equation}
where the subscript 0 identifies the local maximum, and $+$, $-$ identify the adjacent cells.
The values of the $\log(x)$ function are stored in a $10\times 16$ bit LUT.
In the simulation we used a grid step $\Delta=1.6$ mm which is much larger with respect 
to the strip pitch of the sensors of 183 $\mu$m.
The resolutions on the track parameters $x_+$ and $x_-$ 
calculated with the ``artificial retina'' were found to be comparable with the 
 offline results obtained with a $\chi^2$ fit of the clusters to a straight line, 
as shown in Fig.~\ref{fig:resolution}. 
The resolution on the $x_+$ parameter is about 20 $\mu$m for offline reconstructed tracks and 
for tracks reconstructed with the artificial retina. For the $x_-$ parameter we obtain 
a resolution of 27 $\mu$m and 31 $\mu$m with the offline and the artificial 
retina reconstruction, respectively.
Similar results have been obtained with a center-of-mass calculation,
\begin{equation}
\label{eq:CoM_interpol}
x = \frac{\sum_{i=-,0,+}W_ix_i}{\sum_{i=-,0,+}W_i}
\end{equation}
and then applying a correction for the bias, $\delta x = \alpha (x-x_0)$,  
 where $\alpha$ is a constant.
\section{Perspectives for future applications: time-tagged events}
\label{sec:future}
\begin{figure}[!h]
\centering
\includegraphics[width=0.45\textwidth]{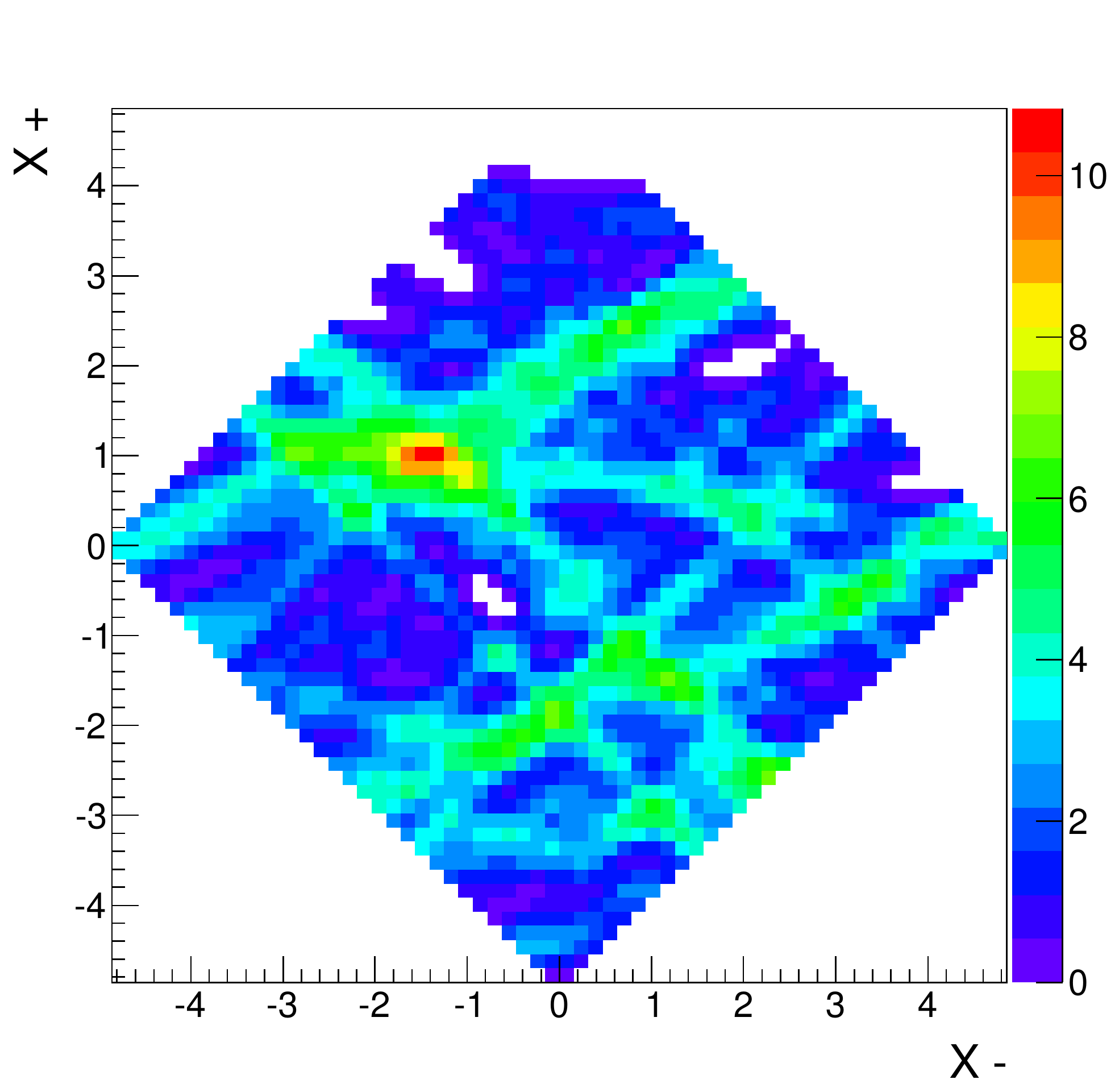}
\includegraphics[width=0.45\textwidth]{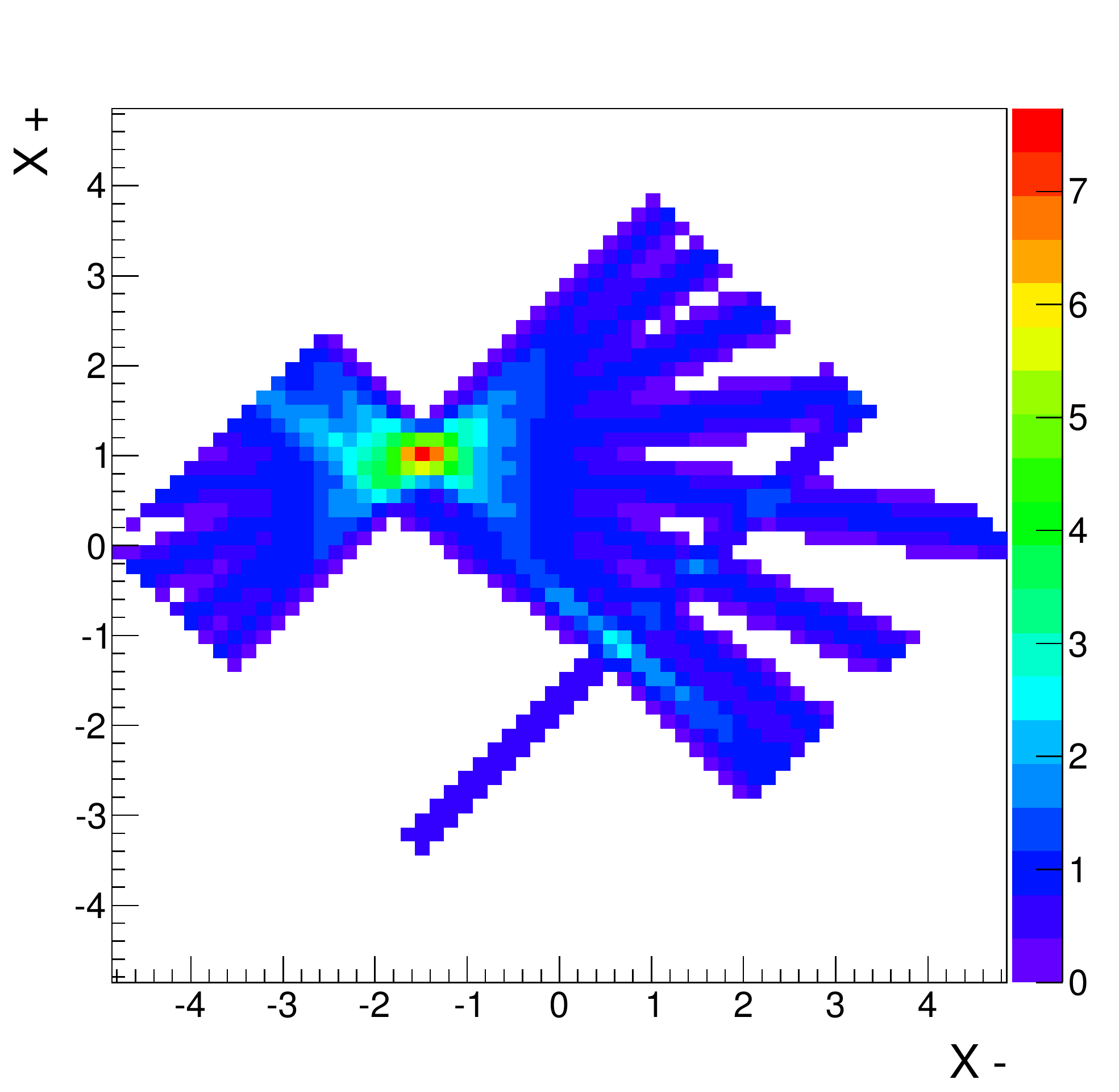}
\caption{The retina response in presence of background hits with no time information (left) and 
with time information right (right).}
\label{fig:track_time}
\end{figure}

Recent advances in silicon detectors make attractive the 
 possibility to include the precise time information of the hit in track reconstruction algorithms.
The gigatracker silicon detector prototypes for the NA62 experiment 
achieved sub-ns time resolution~\cite{Garbolino:2011jx}, and 
recent R\&D activities on ultra fast silicon detectors aim to achieve time 
resolution below 20 ps in the near future~\cite{Cartiglia:2013haa}.
The precise time information of the hit could be used to improve the retina performance. 
In this case, the weight can be defined as a function of the 
 time of the track at the origin, $t$, as 
\begin{equation}
\label{eq:weight_time}
W_{ij}(t) = \sum_k \exp{\left(-\frac{s_{ijk}^2}{2\sigma^2}\right)} \exp{\left(-\frac{t_{ijk}^2}{2\sigma_t^2}\right)}, 
\end{equation}
where $t_{ijk}=|t_k - (t + \Delta t_{ijk})|$ and $t_k$ is the time of the hit at layer $k$, 
$t$ is the time of the track at the origin, and $\Delta t_{ijk}$ is the estimated time for a particle, whose 
 trajectory corresponds to the retina cell $ij$, to arrive at layer $k$. Noise hits out of time are suppressed 
by the corresponding weights where $\sigma_t$ is proportional to the resolution on the time of the hit and 
 can be adjusted for optimal response;
in Fig.~\ref{fig:track_time} it is shown the retina response in presence of background hits
 with no time information (left) and with time information (right).
 The plots correspond to a simulation with detector occupancy of 1\% 
 and resolution on the time of the hit of 100 ps.
 In particular, the time $t$ can be determined from interpolating weights 
 at different times. If the time is divided in 3 intervals around the nominal time of the track, which is 
 supposed to be known with a relative large uncertainty, then the time of the track at the origin can be determined 
 with a precision $\sim \sigma_{h}/\sqrt{N}$, where $\sigma_{h}$ is the 
 resolution on the time of the hit and $N$ is the number of tracking detectors.
As an example, the time of the track $t$ can be determined with a precision of about 30 ps with 8 tracking detectors 
with 100 ps hit time resolution. 

\section{Conclusions}

The design of the first prototype of a tracking system with ``artificial retina'' has been presented. 
It consists of 8 planes of single-sided silicon strip detectors with custom ASIC readout where  
 the ``artificial retina'' architecture is implemented using commercial FPGAs. 
The telescope can operate up to 1 MHz track rate. The working principle can be proved using
 cosmic rays and eventually a test with a beam is envisaged to test the system at higher rates.
 The system is organised in three modules: a switch network for the parallel distribution of the hits, 
 a block of engines for the calculation of the excitation of the cells of the retina 
 and a module for the calculation of the track parameters. The tracking performance of the system are 
 comparable with offline results with a latency of the response $<1 \mu$s which allows fast trigger decisions 
 based on track parameters. 
The system is modular and can be dimensioned for larger tracking detectors with high particle 
rates~\cite{Abba:2014lhcb, Punzi:2014, Tonelli:2014}.

An interesting perspective for future applications is represented by the use of precise information of the time of the hit with 
 a resolution better than 100 ps. This would allow an improved performance of the retina in presence of background hits 
 and also the determination of time-tagged tracks in real time. 


\end{document}